\documentclass[showpacs,twocolumn,aps]{revtex4}
\usepackage[dvips]{graphicx}
\usepackage{amsfonts,amssymb,amsmath,bm}
\usepackage{mathrsfs}

\newcommand{\Real}{\mathop{\rm{Re}}\nolimits}
\begin{document}
\title{Dynamics of wrinkles on a vesicle in external flow}
\author{K.S. Turitsyn and S.S. Vergeles}
\affiliation{Landau Institute for Theoretical Physics,
 Moscow, Kosygina 2, 119334, Russia}
\date{\today}

\begin{abstract}
Recent experiments by Kantsler et. al. (2007) have shown that the relaxational
dynamics of a vesicle in external elongation flow is accompanied by the formation
of wrinkles on a membrane. Motivated by these experiments we present a theory 
describing the dynamics of a wrinkled membrane. Formation of wrinkles is related 
to the dynamical instability induced by negative surface tension of the membrane.
For quasi-spherical vesicles we perform analytical study of the wrinkle structure 
dynamics. We derive the expression for the instability threshold and
identify three stages of the dynamics. The scaling laws for the temporal
evolution of wrinkling wavelength and surface tension are established and
confirmed numerically.
\end{abstract}
\maketitle
Wrinkling of different thin sheets is a well-known effect which can be frequently
observed in everyday life. Usually wrinkle patterns appear due to the
external tensions applied to some material, or as a result of compression of
inextensible films. Main properties of steady and/or equilibrium wrinkling structures
are now well understood \cite{CRM02,CM03}. However, much less is 
known about the dynamics of wrinkle structures. In certain biologically motivated
experiments on membranes and vesicles \cite{SPGFPB06,KSS07}, the wrinkles
are formed because of the instability induced by negative membrane tension,
which is closely related to a buckling instability. Wrinkles formed this way
exhibit non-trivial growth and relaxational dynamics. Theoretical description
of these essentially non-equilibrium processes is a challenging problem which
we attempt to approach in this paper. Although in this letter we focus on the analysis
of the recent experiments \cite{KSS07}, some predictions of our theory are universal
and may be successfully applied to other systems where the formation of wrinkles
is caused by negative tension.

Vesicles exhibit a
variety of different regimes of motion in external fluid flows. These regimes were extensively
studied during the last decade both by the experimentalists \cite{HBVDM97,ALV02,KS05,KS06}
and theoreticians \cite{S99,RBM04,NG04,M06,VG07,NG07,LTV07}. Depending
on the external parameters a vesicle in external shear flow can experience three
types of behavior: tank-treading, tumbling and trembling (also referred as
vacillating-breathing \cite{M06} or swinging \cite{NG07}) \cite{KS06}. Although these
motions correspond to quite non-trivial dynamics, the shape of a vesicle
always remains smooth and can be effectively approximated by an ellipsoid. Recently,
experiments performed by Kantsler et. al. \cite{KSS07} revealed a qualitatively
new effect observed in non-stationary elongation flows. It was shown
that in strong flows the relaxational dynamics of a vesicle is accompanied by
the excitement of high order membrane deformation modes called wrinkles. Such
dynamics can not be described in a framework of low-dimensional models which
were used for the analysis of vesicle tank-treading, tumbling, and trembling. In this
letter we extend these models to include the interaction between the vesicle shape
and the wrinkle structure.

Before proceeding further we would also like to mention the experimental \cite{WRL01}
and theoretical \cite{FS06} investigations of wrinkle formation on microcapsules in external
shear flows. Although this effect is similar to the one discussed here the underlying physics
and main properties of wrinkles are essentially different. For example, the wrinkles which are
observed on vesicles are not stationary and are excited only for a limited amount of time.

This letter is organized as follows: First we discuss the main features of a
vesicle and show that the negative tension leads to instabilities of the flat  membrane.
Second, we present a model of a quasi-spherical vesicle in external flow.
We derive the threshold of the instability and analyze the dynamics of wrinkle
formation in strong flows. We show that one can distinguish three different
stages of the dynamics. The first one lasts for a vanishingly small amount
of time and is characterized by the exponential growth of short wavelength
excitations. During the second stage the amplitude of wrinkles  and their characteristic
wavelengths grow algebraically. We derive the corresponding exponents and
compare them to the experimental results. During the third stage the amplitude
of wrinkles gradually reaches maximum and afterwards decays to zero.  We find the 
scaling estimations for the wrinkle wavelength and the surface tension. In the
end we present the results of our numerical simulations which confirm the analytical
theory. We conclude by discussing the future challenges.

Vesicles are closed lipid bilayers which are incompressible and impermeable to the 
surrounding fluid. These two properties result in the conservation of the vesicle volume 
and the membrane area. Thus any vesicle is characterized by its excess area 
$\Delta$ which is the measure of vesicle's
``nonsphericity'': $A = (4\pi +\Delta) R^2$ where $A$ is the membrane area and $R$ is the
effective vesicle radius, defined by the vesicle volume: $V = 4\pi R^3 /3$.
Free energy of the closed membrane is defined by the Helfrich functional and consists of the
contributions from the bending energy and the surface tension $\sigma$ \cite{Membranes}:
\begin{equation}\label{helfrich}
 F = \int dA \left(\frac{\kappa}{2} H^2 +\sigma\right).
\end{equation}
Here $H$ is the local mean curvature and $\kappa$ is the bending rigidity of a membrane.
Note that for closed membrane geometry, the surface tension $\sigma$ is a quantity
adjusting to other membrane parameters (similar to the pressure in an
incompressible fluid) to ensure a given value of the membrane area
$A$. It is useful to analyze the stability of a flat membrane with a given value of $\sigma$ before
proceeding to the case of a closed membrane with a fixed area. Small perturbations of a
flat membrane can be parameterized by a height function $z=h(x,y)$ which can
be expanded in Fourier harmonics: $h({\bf r}) = \sum h_{\bf k}\exp(i {\bf k r})$. The quadratic part
of the Helfrich energy has the following form:
\begin{equation}
 F = \frac{1}{2}\sum_{{\bf k}} (\kappa k^4 + \sigma k^2)|h_{{\bf k}}|^2.
\end{equation}
For positive $\sigma>0$, this function is minimized by $u = 0$, so the flat membrane state is stable.
However, for negative tension $\sigma <0$, when the membrane is being shrinked, the modes with
$k < \sqrt{|\sigma|/\kappa}$  become unstable.
As we will show, this particular kind of instability is responsible for the formation of wrinkles
in the experiment \cite{KSS07}. More precisely, we will show that the surface tension becomes
negative when the direction of the external flow is reversed instantly. In this case small thermally
induced membrane deformations on top of a smooth vesicle shape are amplified and lead to the
formation of  wrinkles.

In order to study this effect quantitatively we analyze the model of quasi-spherical ($\Delta \ll 1$)
vesicles which proved itself to be very successful in analytical investigations of vesicle dynamics
in external flows \cite{S99,M06,VG07,LTV07}.  The vesicle shape is parameterized  by the
small displacement function $u(\theta,\phi)$: $r = R(1+u)$  which can be expanded in the
series of spherical harmonics:
\begin{equation}
 u(t, \theta,\phi) = \sum_{lm} \left[\frac{2\Delta}{(l-1)(l+2)}\right]^{1/2} u_{lm}(t) {\cal Y}_{lm}(\theta,\phi)
\end{equation}
The external velocity is assumed to have a linear elongational profile with the time-dependent strain:
$\partial_x V_y = \partial_y V_x =
11\sqrt{5}/(16\sqrt{6\pi})\cdot S(t)\sqrt{\Delta}/\tau$, where $\tau = \eta R^3/\kappa$ is the
characteristic time-scale associated with the membrane bending forces. The numerical 
factor in this definition was included to simplify the expressions below.
Throughout the paper we will discuss the experimentally interesting situation when $S(t)$
is the Heaviside-like function $S(t) = - S \mathrm{sign}(t)$.

The dynamical equations describing the dynamics of a quasi-spherical vesicle in
external flow were first derived in \cite{S99} in the leading order in the small parameter
 $\sqrt{\Delta} \ll 1$:
\begin{equation}\label{main1}
        \tau \dot u_{lm} =  S(t) f_{lm} -
        (A_l \sigma + \Gamma_l) u_{lm} + \zeta_{lm}(t),
\end{equation}
Here  $f_{lm} =\delta_{l,2}(\delta_{m,2}+\delta_{m,-2})$ and
$\sigma(t)$ is the dimensionless angularly averaged part of the surface tension
which is a Lagrangian multiplier associated with the excess area conservation constraint 
$\sum |u_{lm}|^2 = 1$.
The numerical coefficients of (\ref{main1}) are given by:
$\Gamma_l = (l-1) l^2 (l+1)^2 (l+2)/(2l+1)(2l^2+2l-1)$, and
$A_l =l(l+1)(l^2+l-2)/(2l+1)(2l^2+2l-1)$. The statistical properties of thermal Langevin forces $\zeta_{lm}(t)$ can be found in \cite{S99}.  In this letter we assume that
the temperature $T$ is small, so that $\zeta_{lm}(t)$ in (\ref{main1}) are negligible during the
evolution of the wrinkle structure. They only affect the dynamics through the
non-zero initial conditions $u_{l,m}(t=0)$ which ensure that the instability
can indeed take the system out of the unstable state. In stationary
planar elongational flows most of the excess area is stored in the ${\cal Y}_{2,\pm2}$ harmonics.
We therefore use the following parametrization : $u_{22} =u_{2,-2}^* = U$ and
$u_{lm} = \sqrt{1-|U|^2}\, w_{lm}$ for $l,m \neq 2,\pm2$.  The fraction of
total excess area stored in the second-order spherical harmonics ${\cal Y}_{2\pm2}$ is given by $|U|^2$.
The argument of $U$ is related to the vesicle orientation angle $\Phi$ in the $xy$-plane:
$U=|U|\exp(-2 i \Phi)$. The normalization condition can be rewritten as
$\sum|w_{lm}|^2 = 1$. The dynamical equations acquire the following form:
\begin{equation} \label{Ueq}
\tau \dot U = S(t) - (A_2\sigma + \Gamma_2) U
\end{equation}
Using the excess area conservation law one can also
find the expression for the surface tension:
\begin{equation}
\sigma = \frac{S(t) \Real U - \Gamma_2 |U|^2 - \bar\Gamma (1-|U|^2)}{A_2 |U|^2 + \bar A (1-|U|^2)}\label{seq}
\end{equation}
where $\bar A = \sum A_l |w_{lm}|^2$ and $\bar \Gamma=\sum \Gamma_l |w_{lm}|^2$. Using
the expression (\ref{seq}) one can easily show that there will be an instability for large enough 
values of $S$. Indeed, for constant positive $S(t)=S$ at $t<0$ the vesicle exhibits small 
thermal fluctuations near the stable point $U=1$. However as $S(t)$ changes sign this state 
becomes unstable and the vesicle starts rotating to the new stable point which corresponds 
to $U = -1$.
The stability of the membrane can be studied by
analyzing the expression (\ref{seq}). The surface tension instantly
becomes negative after changing the velocity field:  $\sigma(t=+0)= -\left(\Gamma_2+S\right)/A_2$.
Negative $\sigma$ can destabilize the high-order harmonics. The explicit condition can be found
from (\ref{main1}). All harmonics of order up to $l$ become unstable if
$A_l \sigma +\Gamma_l <0$, which yields $S > S_l = A_2 \Gamma_l/A_l - \Gamma_2$.
For large $l\gg 1$ one can use the expansions $A_l \sim l/4, \Gamma_l \sim l^3/4$ to
obtain $S_l \sim A_2 l^2 \gg 1$. The most unstable mode can be found
by maximizing the growth increment $-\Gamma_l-A_l\sigma$. One obtains
 $l_0 = \sqrt{S/3 A_2}$ for strong flows with $S \gg 1$.

Below the instability threshold, for relatively small values of $S$ higher order harmonics are
not excited and the dynamics of a vesicle can be well described in terms of a single $U$ variable.
The conservation of excess area  implies that the dynamics is purely rotational and $|U|^2=1$.
The characteristic time-scale associated with the rotational dynamics is estimated as $\tau/S$.

Describing the dynamics of the vesicle above the instability threshold is a considerably more
difficult problem which requires an analysis of a complex nonlinear system (\ref{main1}) with large number
of degrees of  freedom. Fortunately, it is possible to approach this problem analytically for
strong flows which correspond to $l_0 \gg 1$. Although quantitative predictions of such an analysis
can be applied only to strong flows, the qualitative picture is in most aspects the same even for
moderate strains $S \gtrsim 1$.
\begin{figure}[tl]
 \centerline{
 \includegraphics[width=0.4\textwidth]{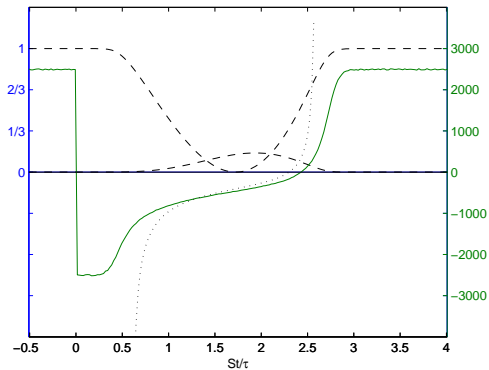} }
 \caption{Temporal dynamics of $|U|^2$, $\Delta_{19}$ (left grid)
 and surface tension $\sigma$ (right grid).
 Dotted line corresponds to the theoretical prediction (\ref{feq}).}
 \label{figure:temp}
\end{figure}

For large values of $S$ one can distinguish between several different stages in the
wrinkle evolution. At the first stage most of the excess area is stored in the
second order angular harmonics and the surface tension can be assumed to be constant
$\sigma = - 2 S/A_2$. Unstable higher order harmonics grow exponentially and the
distribution of $w_{lm}$ becomes centered near the most unstable modes with $l = l_0 \sim \sqrt{S}$, so
the terms $\bar A,\bar \Gamma$ can be estimated as $\bar A \sim \sqrt{S}$ and $\bar\Gamma\sim S^{3/2}$.
This rapid growth saturates when the total excess area, stored in higher-order harmonics, becomes large enough
so that their contribution to the surface tension becomes comparable to the contribution from the external flow.
Formal condition can be found from (\ref{seq}): $(1-|U|^2) \sim S^{-1/2} \ll 1$. Note that at this time most
of the excess area is still stored in the second-order harmonics. The characteristic
duration of this stage can be estimated as $\tau S^{-3/2}$.
This time is much less than the characteristic relaxation time of the variable $U$,
which can be estimated as $\tau S^{-1}$ from the equation (\ref{Ueq}) as in the situation below
the instability threshold.

\begin{figure}[tr]
 \centerline{
 \includegraphics[width=0.4\textwidth]{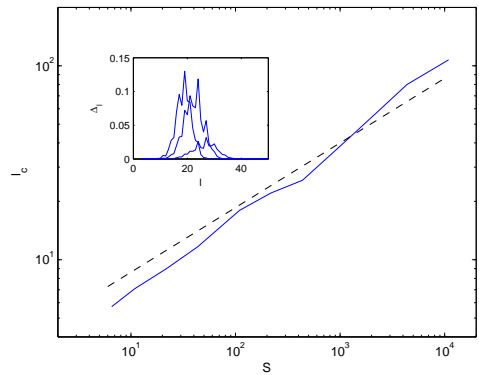} }
 \caption{ $l_*$ versus the strain $S$. Linear curve corresponds to the scaling $l_* \propto S^{1/3}$.
 Inset: Distribution of $\Delta_l$ as a function of $l$ at the moments $tS/\tau = .7:1.3:2.2$ (right to left).}
 \label{figure:av_l}
\end{figure}

Second stage starts at $t\gtrsim \tau S^{-3/2}$ after the exponential growth has saturated.
During the second stage higher order harmonics $w_{lm}$ provide the main contribution
to the surface tension (\ref{seq}) which can be approximated as $\sigma = -(\bar\Gamma/\bar A)$. Note that it
depends only on  the distribution of excess area over the higher-order harmonics $w_{lm}$. In order
to find this distribution for large $l \gg 1$ one can use the formal solution of (\ref{main1}):
$w_{lm}(t) = C w_{lm}(0) \exp\left(-\Gamma_l t/\tau + A_l \rho\right)$ where $C$ is the normalization
constant and $\rho = -\int_0^t dt' \sigma(t')/\tau$.   The distribution of excess area
$\Delta_l = (1-|U|^2)\sum_m |w_{lm}|^2$
is a narrow function of $l$  centered around some $l_*$ which is determined as maximum
of $A_l \rho - \Gamma_l t/\tau$: $l_* = \sqrt{\rho\tau/3 t} \sim \sqrt{S}$. The characteristic width of the distribution
can be also estimated in a usual way: $\delta l \sim (l_* t/\tau)^{-1/2}$. It is much less than $l_*$ for
$t \gg \tau S^{-3/2}$. Narrowness of the distribution allows one to find the exact expression for the surface
tension: $\sigma= -\bar \Gamma/\bar A = - l_c^2 = -\rho\tau/3t$. Using the definition of $\rho$ we
obtain the following closed equation:
\begin{equation}
 \dot \rho = \rho/3t
\end{equation}
One can find its solution using the initial condition for the surface tension:
$\sigma \sim \sqrt{S}$ at $t \sim \tau S^{-3/2}$.
This yields  $\rho = c (t / \tau)^{1/3}$, where $c$ is a constant of order
unity. Similarly we obtain $\sigma = - (c/3)  (t/ \tau)^{-2/3}$ and $l_* = \sqrt{c/3} (t/\tau)^{-1/3}$.
Therefore,  second stage of the dynamics is characterized by the algebraic decay of the
surface tension and by the narrow spectral distribution $\Delta_l$ of the wrinkling structure, whose
peak smoothly drifts towards small  $l$. As the absolute value of $l_*$ goes down the external
velocity contribution $S(t) \Real U$ to the surface tension in (\ref{seq}) becomes important
again. Comparing different contributions in (\ref{seq}) one can estimate the duration of the second
stage as $\tau/S$. In the end of the second stage the peak of the $\Delta_l$ distribution is centered near
$l_* \sim S^{1/3}$. This scaling law  relates wavelength of wrinkles to the strain $\lambda = R S^{-1/3}$
and is one of  the main results of this letter. Note that this wavelength is much smaller compared
to the wavelength of the initially most unstable mode which is given by $R S^{-1/2}$. Our scaling is
in reasonable agreement with the experimental results of Kantsler et. al. \cite{KSS07}.

Second stage is followed by third one during which the surface tension in (\ref{seq}) is determined
both by the external velocity and the wrinkles. The characteristic value of the surface tension can be
estimated as $S^{2/3}$ which is much smaller than its initial value $\sigma(0)\sim S$.  Due to a significant
decrease of the surface tension, the dynamics of the second-order harmonics during the second stage
is determined solely by the  external velocity term $S(t)$ in (\ref{Ueq}). Therefore one can find the solution
of (\ref{Ueq}) in the leading order: $U(t) = 1 - S t/\tau$. The characteristic amplitude of wrinkles obeys the
following simple law:  $\sqrt{1-|U|^2} = \sqrt{(2- S t/\tau)St /\tau} $. In order to analyze the evolution of the surface
tension $\sigma$ and the wrinkles wavelength $\lambda \sim R/l_*$ one has to solve the equation
$\tau \dot \rho = \sigma$ keeping all the contributions to the surface tension. With the parametrization
$\rho(t) = S^{-1/3} f(St/\tau)$ the problem is reduced to the ordinary differential equation:
\begin{equation}\label{feq}
 \frac{d f(x)}{d x} = \frac{f(x)}{3 x} +4 \frac{1-x}{2-x}\sqrt{\frac{3}{x f(x)}}
\end{equation}
with the initial condition $x^{-1/3} f(x) = c$ for $x = S^{-1/2}$ determined by the second 
stage. It is possible to solve (\ref{feq}) analytically: 
$f(x) = x^{1/3} \left[\log C \sqrt{S} x (2-x)\right]^{2/3}$, where $C$ is a constant of order unity. 
Note, that this solution adds only logarithmical factors of order $\log S$ to the previously
derived scalings $l_* \sim S^{1/3}$ and $\sigma \sim S^{2/3}$. 
Although the second stage is formally present in the solution of (\ref{seq}) and
corresponds to the limit $x\ll 1$ we purposely separate it from the third one because
of its universality: during the second stage the surface tension is determined solely
by the wrinkle structure. Therefore one may expect that the scaling laws
$l_* \propto t^{-1/3}$ and $\sigma \propto t^{-2/3}$ will also hold for other problems
where the initial negative surface tension was caused by different means.

We have tested our results by direct numerical simulations of equation (\ref{main1}).
The total number of  harmonics in our simulations was $l_{max} \approx 2 \sqrt{S}$. 
The temperature which determines the power of the
thermal noises $\zeta_{lm}$ in (\ref{main1}) was taken to be $T = 10^{-4}\kappa$. On
FIG. \ref{figure:temp} one can see the results of simulations for $S=10^3$. Three
temporal stages with qualitatively different dynamics are clearly distinguishable. 
The behavior of $|U|^2$ is very close to quadratic and the dynamics
of $\sigma(t)$ may be well fitted by the solution of (\ref{feq}). The scaling law
$l_* \propto S^{1/3}$ was confimed by simulations at different $S$ (FIG. \ref{figure:av_l} )

In conclusion we compile a list of main results presented in this paper. Motivated by recent
experiments \cite{KSS07}, we studied the relaxational dynamics of a vesicle in an
elongational flow. We showed that high order membrane deformation modes
are excited by the negative surface tension induced by external flow.
For quasi-spherical vesicles we have found an analytical expression for the instability
threshold and analyzed the evolution of the wrinkle structure. We identified three
stages of the dynamics. The first stage corresponds to $t \lesssim \tau S^{-3/2}$ and
is characterized by the exponential growth of unstable high order harmonics with the
characteristic scales of order $\lambda \sim R S^{-1/2}$.
This rapid growth quickly saturates and is followed by the second stage. For $\tau S^{-3/2}\ll t \lesssim \tau/S$
the surface tension decays algebraically as $\sigma(t) \propto t^{-1/3}$ and the characteristic
wavelength of wrinkles grows as $\lambda \propto t^{1/3}$. Characteristic amplitude of the
wrinkles grows as $\sqrt{t}$. During the third stage  which ends at $t = 2\tau/S$ wrinkle
amplitude behaves like $\sqrt{(2-S t/\tau)S t/\tau}$ and the characteristic wavelength
can be estimated as $\lambda \sim R S^{-1/3}$.

Finally we would like to note, that the algebraic growth of the characteristic wrinkle wavelength 
although with different exponent $\lambda(t)\propto t^{1/4}$ was observed in the studies 
of the thin rods dynamics experiencing the buckling instability \cite{GMP98}. We believe that
the algebraic growth of wrinkle wavelength is a universal property of the interface dynamics
experiencing the buckling instability. Although in this letter we considered a particular
experiment, we believe that the scaling laws for the wrinkle wavelength $\lambda(t)$ and
for the surface tension $\sigma(t)$ derived for the second stage are universal and can be
observed in other experiments where the total area of the interface is conserved and the wrinkle
growth is initiated by the external forces leading to negative surface tension.

We are indebted to V. Kantsler and V. Steinberg for turning our attention
to this problem. We also appreciate the fruitful discussions with V. Lebedev.
This work was supported by the RFBR grant and by Dynasty and
RSSF foundations.

\end{document}